\documentclass[10pt]{article}

\usepackage{iclr2020_conference, times}
\usepackage{epsfig}
\usepackage{graphicx}
\usepackage{amsmath}
\usepackage{amssymb}
\usepackage{natbib}
\usepackage{float}
\usepackage{url}
\usepackage{booktabs}
\usepackage{subfigure}
\usepackage{caption}
\usepackage{enumitem}
\usepackage[margin=1.3in]{geometry}
\usepackage{layout}

\usepackage[breaklinks=true,bookmarks=false]{hyperref}

\iclrfinalcopy

\begin{document}

\title{\vspace{-3.5cm}Incrementally Improving Graph WaveNet Performance on Traffic Prediction}

\author{
Sam Shleifer\\
Stanford University\\
{\tt\small shleifer@stanford.edu}
\and
Clara McCreery\\
Stanford University \\
{\tt\small 
mccreery@stanford.edu}
\and
Vamsi Chitters\\
Stanford University\\
{\tt\small vamsikc@stanford.edu}
}
\maketitle

\begin{abstract}
    We present a series of modifications which improve upon Graph WaveNet's previously state-of-the-art performance on the METR-LA traffic prediction task. The goal of this task is to predict the future speed of traffic at each sensor in a network using the past hour of sensor readings. Graph WaveNet (GWN) is a spatio-temporal graph neural network which interleaves graph convolution to aggregate information from nearby sensors and dilated convolutions to aggregate information from the past. We improve GWN by (1) using better hyperparameters, (2) adding connections that allow larger gradients to flow back to the early convolutional layers, and (3) pretraining on an easier short-term traffic prediction task. These modifications reduce the mean absolute error by .06 on the METR-LA task, nearly equal to GWN's improvement over its predecessor. These improvements generalize to the PEMS-BAY dataset,  with similar relative magnitude. We also show that ensembling separate models for short-and long-term predictions further improves performance. Code is available at \url{https://github.com/sshleifer/Graph-WaveNet}.
\end{abstract}

\section{Introduction}
Americans spend an average of 51 minutes per day driving, and 11 minutes sitting in traffic.\footnote{2019 Urban Mobility Report, \url{https://static.tti.tamu.edu/tti.tamu.edu/documents/mobility-report-2019.pdf}} Improved traffic prediction can help minimize traffic congestion by warning travelers of delays, so they can adjust their routes or departure times. Furthermore, the insights gained from improved spatio-temporal modeling might be extended to other important applications, like ecology, epidemiology, and climatology.

This paper focuses on traffic forecasting: predicting the future traffic speeds at each sensor in a network given recent traffic speeds at each sensor and spatial information. The road network is represented as an adjacency matrix containing the on-road distance between sensors.

\section{Dataset}

We use the same traffic dataset as \textit{Graph WaveNet} and its predecessor, \textit{DCRNN}, which we discuss further in our Related Works section:\\\\
\includegraphics[scale=0.3]{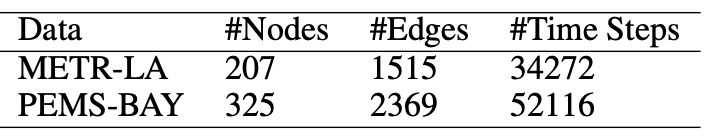} \\\\
\textbf{METR-LA}: This dataset comes from sensors along the Los Angeles County highways which record the velocities of passing vehicles. Each sensor's readings are binned into 5-minute chunks and subsequently averaged. We ran all experiments on the METR-LA dataset, and then verified that our modifications improve accuracy on a larger but similarly structured dataset, PEMS-BAY, which contains 325 sensors and 6 months of data from the bay area.
\begin{table}[H]
    \begin{minipage}{.5\linewidth}
    \centering
    \begin{figure}[H]
        \centering
    \includegraphics[scale=0.25]{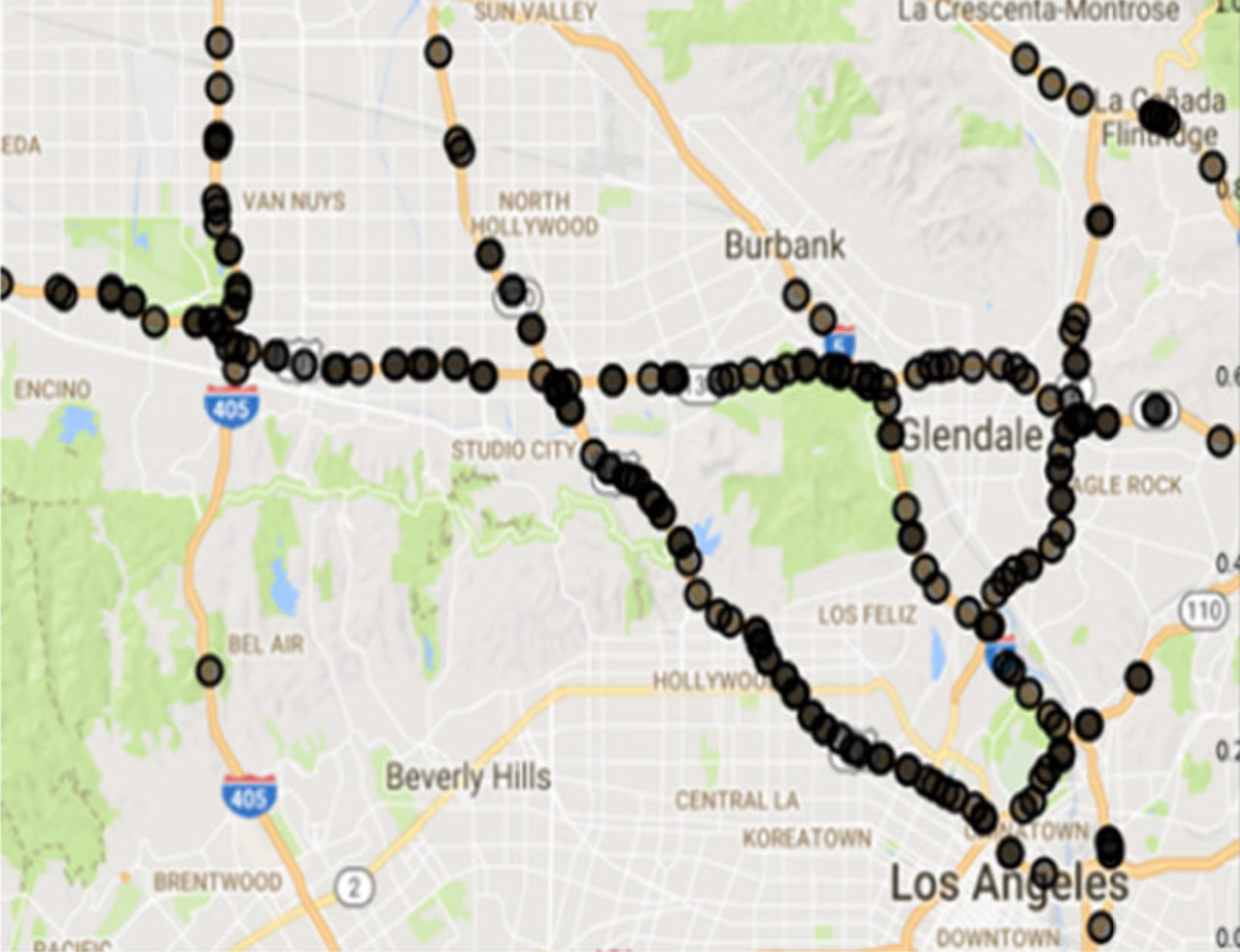}
    \caption{METR-LA Sensor distribution}
    \end{figure}
    \end{minipage}
    \begin{minipage}{.5\linewidth}
      \centering
        \begin{figure}[H]
        \includegraphics[scale=0.41]{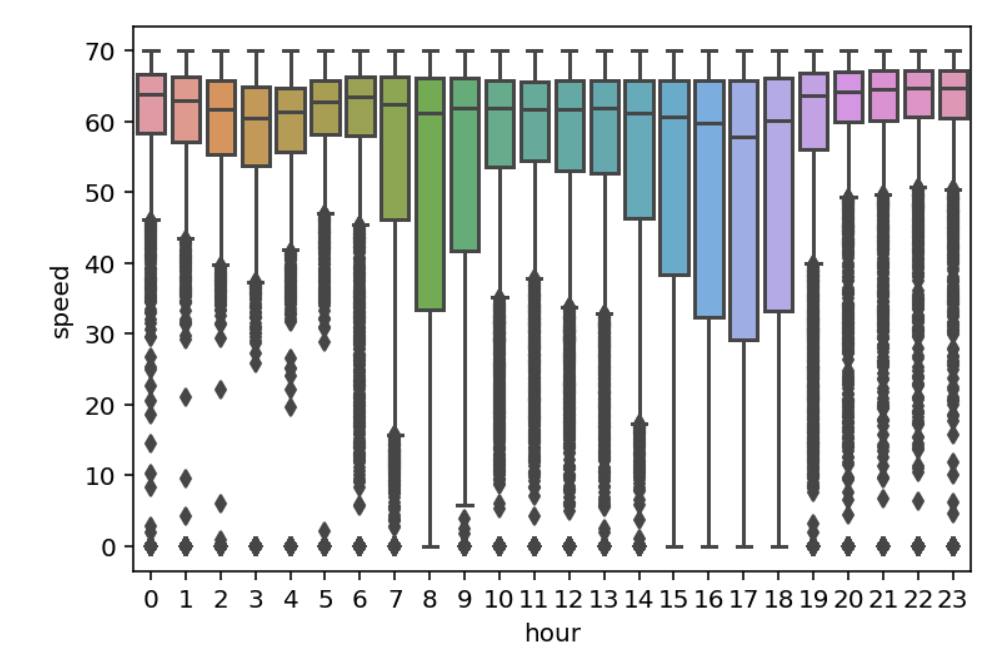}
        \caption{Speed vs. time of day in LA across all sensors. T raffic slows down around morning and evening rush hours.}
        \end{figure}
    \end{minipage}
    \label{tab:degree_stats}
\end{table}

The dataset also includes a directed adjacency matrix that reflects the connection strength based on the driving distance from one sensor to another in meters. Specifically:
\[W_{ij} = max(e^{-\frac{D_{ij}}{\sigma}^2} - k, 0)\]
where $W_{ij}$ is the edge weight between nodes $i$ and $j$, and $D_{ij}$ is the driving distance. A normalizing term, $\sigma$, is the standard deviation of all of the distances between any two nodes in the graph (standard deviation of $D$), and $k$ is some threshold below which all weights are set equal to 0. All self-loops are included.  In the DCRNN and Graph WaveNet papers, $k = 0.1$, which we use as well.

The features for each node at a given timestep have shape \texttt{(2, 12)} corresponding to speed and time-of-day for each of the 12 timesteps preceding the measurement of interest. Since the model needs to see all nodes in the network to perform graph convolution, batches are shaped \texttt{(BatchSize=64, NumNodes, 12, 2)}.

\section{Evaluation metric}
Similar to the WaveNet \citep{WaveNet} paper, we use Mean Absolute Error (MAE) as both the loss that the model back-propagates and the reported metric:
$$MAE_{t} = \frac{1}{Z} \sum_{i \in Sensors} |PredictedSpeed_{i,t} -
Speed_{i, t}|$$
where $Z = |Sensors|$ is a normalizing constant, and $t$ is the prediction horizon. We also continue the friendly convention of assigning 0 loss to the roughly 5\% of observations where $Speed_{i, t}=0$, which, according to the DCRNN authors, represent intervals where no cars passed over the sensor, rather than standstill traffic. The metric MeanMAE
$$MeanMAE = \frac{1}{12} \sum_{1 \leq t \leq 12}  MAE_{t}$$ represents the average MAE across all prediction horizons. Note that each $t$ is a 5 minute increment so the horizon $t=12$ is one hour in the future.

\section{Related Work}
There are two recent papers in the traffic-prediction lineage whose work we leverage heavily. DCRNN \citep{li2018dcrnn_traffic} is the first paper to use graph convolution on the road network graph, which they split into two adjacency matrices: $D^{-1}_OW$ for the downstream traffic and  ($D^{-1}_IW^T$) for the upstream traffic. 
They then compute a \textit{diffusion convolution operation}, shown in equation (2).  Graph WaveNet adopts this split, and we do not modify it.

DCRNN uses a Gated Recurrent Unit (GRU) to model short-term temporal interactions, and significantly outperforms all predecessors, which are either non-neural or do not take full advantage of the network structure.
Graph WaveNet \citep{gwn} addresses two shortcomings in DCRNN.
First, the distance-based adjacency matrix built by DCRNN assumes that sensors are always correlated to one another when they are spatially close. This is a much better prior than training from a random adjacency matrix, but may overemphasize the importance of an edge between two nearby sensors even if the route between them is unlikely to be traversed. For example, if two sensors are on nearby one-way roads going in opposite directions, both $E_{i,j}$ and $E_{j,i}$ would be large in the adjacency matrix, even though traffic at $j$ is not very predictive of traffic at $i$ and vice versa.\\\\
Graph WaveNet \citep{gwn} attempts to relax this assumption by learning source and target embeddings of size $d=10$ for each sensor, and using:\\
\[ W_{learned} = SoftMax(ReLU(matmul(emb_{src}, emb_{dest}))) \]
\newline in addition to the fixed, distance-based adjacency matrices.

\begin{figure}[H]
\centering
    \includegraphics[scale=0.15]{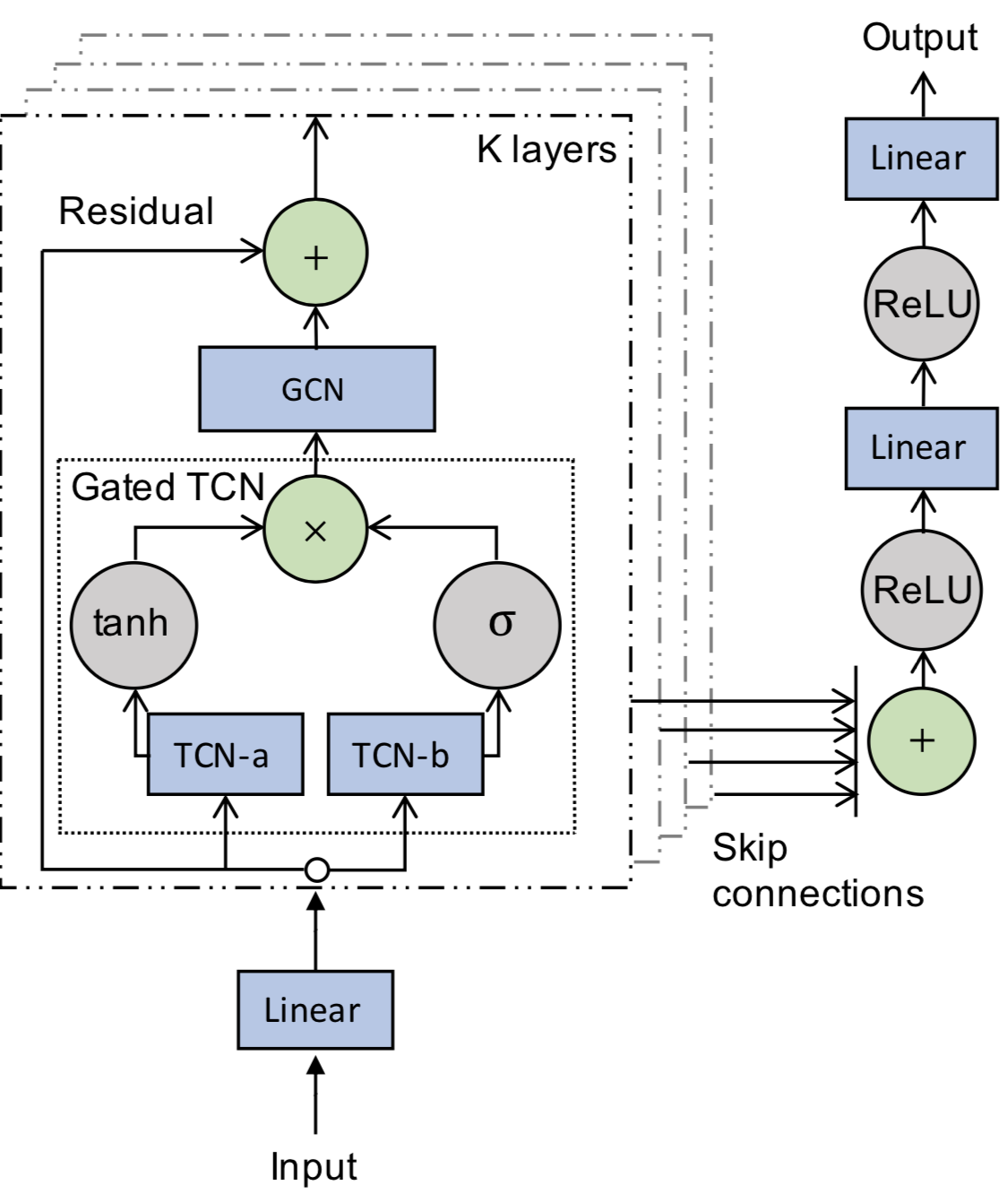}
    \caption{GWN's diagram of their WaveNet inspired blocks. Architecture adopted from Graph WaveNet. The Gated TCN is a 1D temporal convolutional module, while the GCN unit incorporates the learned adjacency matrix.
    }
    \label{gwn_architecture}
\end{figure}

DCRNN also used a very expensive encoder decoder setup, which took a long time to train. GWN replaces the GRU based encoder decoder setup with blocks inspired by the original WaveNet \citep{WaveNet}, which was first used for audio generation. These blocks, shown in Figure \ref{gwn_architecture} use 1D and 2D dilated convolutions, and predict all 12 timesteps at once, rather than decoding one step at a time. Put together, these changes reduce training time by a factor of 6 and inference time by a factor of 10. They also help performance: GWN improves over DCRNN in MeanMAE by .07 (from 3.11 to 3.04).

\section{Improvements}

\subsection{Hyperparameters}

We adjust several hyperparameters across all of our models and observe a drop in MAE of 0.03 compared to that reported by the original Graph WaveNet. The impact of each individual hyperparameter change is shown in Table \ref{mods}.

\textbf{Learning Rate Decay:} We multiply the learning rate by 0.97 after each epoch. Graph WaveNet did not use learning rate decay, effectively multiplying by 1 after each epoch.

\textbf{Number of Filters:} We find that increasing the number of filters used by all layers in the Gated-TCN block and the GCN block from 32 to 40 improves performance with just a 5\% increase in training time. This change increases the number of trainable parameters from 309,400 to 477,872 (54\%).

\textbf{Gradient Clipping:} We find that clipping gradients to $L2 = 3$ provides a lower error than the original Graph WaveNet, which clipped gradients to $L2 = 5$. 

\textbf{Missing Data Representation:} As previously mentioned, the loss function assigns 0 loss to observations where the sensor reading is 0, which means no vehicles passed the sensor during the 5 minute interval. Roughly 5\% of the data contain these zeroes, and although they do not cause loss when they are in the targets,  they still contribute nonsensical numbers to the predictions of nearby sensors when they are in the features. With this representation,  the model must learn that lower speed measurements indicate higher traffic, except when the speed measurement is 0, in which case it represents no traffic. We therefore replace these 0's with the average speed in the training data, and get another .01 improvement in MAE, as well as faster convergence. 

\subsection{More Skip Connections}
From Figure 4, we note that the output of the Gated TCN at each layer passes through a GCN directly, and intermediate results are not carried forward independent of that. As a result, during backpropagation, Graph WaveNet passes very small gradients back to the early convolution layers that are furthest from the loss calculation. To rectify this, in each block, we add the output of the Gated TCN to the output of the GCN at each block, effectively providing a skip connection around the GCN, similar to the skip connections that already exist between entire layers. To rephrase, where each layer previously computed 
\[x_{i+1} = GraphConv(x_i),\]
we now compute \[x_{i+1} = x_i + GraphConv(x_i),\] thereby allowing the gradients a more direct route to the earliest layers in the network. This change makes the gradients into the early blocks larger, reinforcing the need to clip gradients more restrictively. In our model gradients are clipped at $L2_{norm}= 3$, a reduction from $L2_{norm}= 5$ in the original GWN.

\subsection{Pretraining on shorter prediction horizons}

\begin{figure}[h]
    \centering
\includegraphics[scale=0.6]{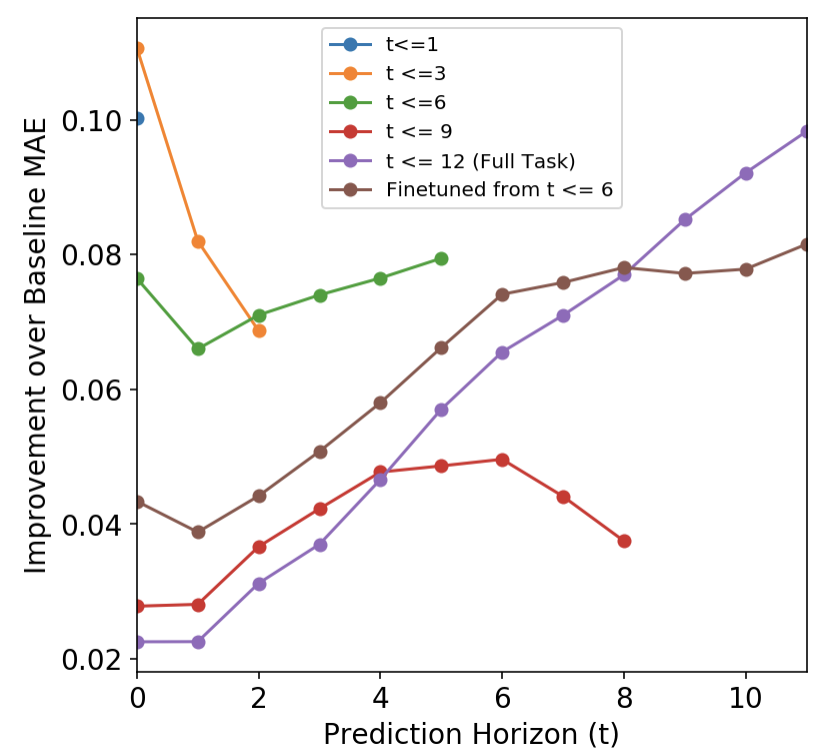}
\caption{MAE improvement over baseline for different pretraining and finetuning tasks. For example, $t \leq 3$ is a model trained to predict 3 timesteps (5, 10, and 15 mins) ahead, and therefore only has 3 data points in the chart. We observe that models trained to predict short-range traffic do so better than models trained to predict both short- and long-range. We plot improvement over baseline rather than absolute MAE because the MAE at later timesteps increases significantly, making it difficult to visualize the differences between models.}
\label{transfer_learning}
\end{figure}

We find that models trained on a short-term subset of the full 1 hour prediction horizon perform better on these short-term predictions than models trained to predict the full hour, as shown in Figure \ref{transfer_learning}. In other words, models that are allowed to specialize on the short-term traffic do better for short-term predictions than those which are trying to learn short- and long-term patterns simultaneously.

This finding motivated us to initialize the weights for the full task with weights learned on shorter range tasks, providing a .01 reduction in MAE (averaged across 2 runs). This pretrain/finetune approach takes only 7 minutes more than training from scratch, because the short-horizon model converges after 60 epochs, and the finetuned model converges after only 29 epochs. Training on the full task from scratch converges after 100 epochs.\footnote{80 minutes on an NVIDIA v100 GPU. And, somehow, 4 hours on a T4 GPU!}

As the model improves on longer horizons, it worsens on shorter horizons. In Figure \ref{transfer_learning}, the brown curve underperforms the green curve (from which it was finetuned) at short horizons. For this reason, using an ensemble of the green short-horizon model (for short horizon predictions) and the brown finetuned model (for longer horizon predictions) achieives a further .01 reduction in MeanMAE with no additional training cost. We did not spend time optimizing hyperparameters for finetuning or ensembling and there may be modifications that reduce the ``forgetting" of the short-term pretraining task.

We hypothesize that shorter prediction ranges allow the model to ignore further away nodes, and focus on a simpler task, but not all of our experimental data supports this. For example, a model trained to only predict $7 \leq t \leq 12$ (not shown in plot) performed much worse than a model trained on the full task: $1 \leq t \leq 12$, even though it should be able to "focus" more. Second, a model trained to predict only the next 45 minutes (red) performed worse on validation data than a model trained to predict the next hour (purple). These dynamics merit further investigation.

\section{Results}

\subsection{Overview}
Using a combination of small modifications to the Graph WaveNet architecture and training schedule, we reduced this model's error on the METR-LA task from 3.04 down to 2.98, a new state-of-the-art on this traffic prediction task. An overview of the improvements achieved by each of the modifications can be found in Table~\ref{results}.

Following the precedent set by DCRNN and Graph WaveNet, we split the data into 70\%, 10\%, and 20\% for our train, validation, and test splits respectively. The splits reflect time: the train set precedes the validation set which precedes the test set. We considered only the validation MAE for model selection and early stopping, but we report test MAE in tables and figures to facilitate comparison with the GWN paper, which does not report validation results. Across all of our runs, test set MAE was 94\% correlated with validation MAE, and the improvements we discuss improved both validation MAE and test MAE. After deciding on the final set of approaches, we ran everything on PEMS-BAY to verify that our changes generalize to another traffic dataset.

\begin{table}[!htbp]
\centering
\begin{tabular}{@{}lllll|llll@{}}
\toprule
METR-LA              &         &         &         &         & PEMS BAY        &         &         &          \\ \midrule
Model                & 15 min & 30 min & 60 min & Mean & 15 min & 30 min & 60 min & Mean \\
 & & & & MAE & & & & MAE \\
\midrule
DCRNN (Reported)     & 2.77    & 3.15    & 3.60    & 3.11     & 1.380   & 1.740   & 2.070   & 1.680    \\
GWN (Reported)       & 2.69    & 3.07    & 3.53    & 3.04     & 1.30    & 1.63    & 1.95    & 1.58     \\
                     &         &         &         &          &         &         &         &          \\
GWN (Replicated)     & 2.70    & 3.10    & 3.55    & 3.06     & 1.316   & 1.647   & 1.968   & 1.591    \\
                     &         &         &         &          &         &         &         &          \\
GWNV2 (Ours)  & 2.67    & 3.04    & 3.45    & 3.00     & 1.303   & 1.622   & 1.909   & 1.560    \\
Finetuned (Ours)     & 2.66    & \textbf{3.03}    & \textbf{3.47}    & 2.99     & 1.299   & 1.613   & \textbf{1.896}   & 1.552    \\
RangeEnsemble (Ours) & \textbf{2.64  }  &\textbf{3.03}    & \textbf{3.47}    &\textbf{2.98  }   & \textbf{1.286}   & \textbf{1.606}   & \textbf{1.896}   & \textbf{1.546}    \\ \bottomrule
\end{tabular}
\caption{MAE for different models. GWNV2 incorporates the skip connection around GCN as well as modified hyperparameters (further breakdown of how each modification contributes to this number can be found in Table \ref{mods}). Finetuned model is initialized with weights of a model trained to predict traffic up to 30 minutes forward. RangeEnsemble uses that same 30 minute model to predict the first 30 minutes, and the finetuned model to predict minutes 35-60. All reported metrics are the average of 2 runs.}
\label{results}
\end{table}

\section{Analysis}
\subsection{Failed Experiments}
To save other researchers cloud computing credits and partially offset our carbon impact, we briefly list experiments we tried that did not improve validation MAE. 

\begin{enumerate}
    \item Adding day of week information.
    \begin{itemize}
        \item Empirically, traffic speeds are about 10 mph slower on weekdays at 5pm than on weekends. However, adding the day of week to the model, either in a one hot representation, or as a scalar, did not influence the metrics, or even make the model converge faster. Either this information needs to be added in a different format, or this indicates that short-term traffic readings already capture these trends adequately. Intuitively, it seems like there should be a few minutes, right before rush hour starts, where there is very little recent traffic, and knowing the day of the week would help predict whether there will be a rush hour. It could also make sense to use a business calendar to build a binary isWorkday feature.
    \end{itemize}
    \item Using the last 75 minutes instead of the last 60.
    \item Changing dropout, learning rate, weight decay, activation functions.
    \item Cyclical learning rates.
    \begin{itemize}
        \item Cyclical learning rates \citep{Smith_2017} lead to faster training and better metrics in many contexts, but not this one!
    \end{itemize}
    \item Transfer learning between road networks
    \begin{itemize}
    \item We tried finetuning a model on METR-LA based on a fully trained checkpoint on PEMS-BAY (besides the adjacency matrices which are the wrong shape, and also obviously shouldn't transfer). This model was about .03 MAE worse than training from scratch, although it trained faster, suggesting some local minimum was reached. This also isn't very useful practically since there are only two such traffic datasets that we know of.
    \end{itemize}
    \item Larger batch sizes and half precision.
    \begin{itemize}
    \item Switching to half precision actually made training a tiny amount \textbf{slower}, and hurt metrics considerably. At default batch size, the model is not memory bottlenecked, requiring only 4GB of GPU RAM. We tried to increase batch size both with and without half precision. In both contexts, larger batch size impacted performance negatively, (even with commensurate learning rate increases), and multiplying batch size by 8 only sped up training by 10\%.  
    \end{itemize}
    \item Using transformers instead of 1D Convolutions.
    \begin{itemize}
        \item  Passing  the features through 1 transformer with 1 attention head at the very beginning of the model hurt metrics. Adding any transformer inside the block, of which there are 8, made training unbearably slow. We did let a run finish and it was very bad, but too slow and not promising enough to iterate further.
    \end{itemize}
    \item Removing the information deletion in the block. 
    \begin{itemize}
        \item Each layer downsamples the time dimension of the tensor by 1 or 2, so skip connections of previous layers discard the outputs for the ``extra timesteps" from the earliest part of the previous output.  We tried to address this by mean-pooling the channels that would have been discarded. This did not improve performance, and reflects the trend that the most recent 30 minutes is much more useful for prediction than the next 30.
    \end{itemize}
    \item Initializing the learned adjacency matrix from the SVD decomposition of the provided road network.
    \item Directly backpropagating RMSE instead of MAE improves RMSE slightly but hurts MAE.
    \item Changing the temperature of the softmax applied to the learned adjacency matrix.
    \item Changing number of layers per block. (This slows things down a lot!)
    \item Reducing/increasing the size of the learned node embeddings, $emb_{src}$ and $emb_{dest}$.
    \item Adding batch normalization after the graph convolution step.
\end{enumerate}

\subsection{Ablations}

In Table \ref{mods}, we show that none of our modifications can be removed architecture without impacting performance. In particular, learning rate decay, a simple change, is essential.
\begin{table}[!htbp]
\centering
\begin{tabular}{@{}ll@{}}
\toprule
Modification          & Mean MAE \\ \midrule
GWN baseline (no modifications)              & 3.057    \\
without n channels=40          & 3.024    \\
without skip connection & 3.013    \\
without 0 replacement & 3.010  \\
without grad clipping=3        & 3.023    \\
without lr decay      & 3.032    \\
with all modifications            & \textbf{3.002} 
\\ \bottomrule
\end{tabular}
\caption{We experiment with removing each of our proposed modifications from the final architecture, and resetting them to default values. Default values: clip=5, lr decay=1 (no decay), n channels=32.}
\label{mods}
\end{table}

We also provide an ablation over the number of timesteps used. We find that using more history to make predictions does tend to help, but the improvement is not monotonic, and starts to plateau after 5 time steps.  In the table below, we show the performance of models that are given access to different amounts of history. All prediction horizons are 60 minutes (12 timesteps) into the future.

\begin{table}[H]
\centering
\begin{tabular}{@{}lllllllll@{}}
\toprule
History Length & 1     & 2     & 3     & 4     & 5     & 6     & 9     & 12  \\ \midrule
MAE            & 3.058 & 3.034 & 3.014 & 3.017 & 3.007 & 3.015 & 3.024 & 3.002 \\ \bottomrule
\end{tabular}
\caption{Mean MAE of models given access to different amounts of history. For example History Length=6 uses the last 30 minutes. All models are trained to predict the next hour.}
\label{history}
\end{table}

Finally, we verify that graph convolution in general and Graph WaveNet's learned adjacency matrix are crucial for strong performance. Without any graph convolution (or other modifications), we get Mean MAE of 3.59. Without the learned adjacency matrix, we get mean MAE of 3.08. And, without time of day information or modifications, the model achieves MeanMAE of 3.1. This last result is surprisingly strong, and supports the overall trend we observed that short-term traffic information is far more useful than longer-term information and attributes like time-of-day, day-of-the-week, or traffic readings from more than an hour ago.

\section{Conclusion and Future Work}
We showed that a few small modifications to the Graph WaveNet architecture: increasing the size of some internal convolutional layers, a skip connection, learning rate decay, and changing the representation of missing data, improves performance on both of the traffic datasets with which we experimented. Replacing 0s in the features with the average speed in improved results, but we believe that slightly more complex interpolation schemes, like copying the most recent nonzero sensor reading may produce further gains.

We also show that pretraining on easier short-term traffic speeds improves full task performance, but that as the finetuned model learns longer horizon relationships, it loses some of its performance on the shorter horizons. Mysteriously, a model trained to predict only longer-term traffic horizons performed poorly. Future work should aim to solve this mystery and ideally find a way to train one model that excels at both short- and long-term traffic prediction.
\newpage
\bibliography{gwnv2}
\bibliographystyle{iclr2020_conference}
\end{document}